\documentclass[aps,prl,twocolumn,showpacs,nofootinbib]{revtex4-1}
\usepackage[dvipsnames]{xcolor}
\RequirePackage[T1]{fontenc}
\RequirePackage{mathptmx}
\usepackage{fullpage}
\usepackage{slashed}
\usepackage{graphics,bm}
\usepackage{epsfig}
\usepackage{graphicx}
\usepackage{amsmath}
\usepackage{amssymb}
\usepackage{bbold}
\usepackage{wrapfig}
\usepackage{color}
\usepackage{hyperref}
\usepackage{mathtools}
\usepackage{cancel}
\usepackage{multirow}
\usepackage{upgreek}
    \usepackage[vertfit]{breakurl} 

\DeclareMathOperator{\Tr}{Tr}
\begin{document}
\title{Topological Susceptibility in a Uniform Magnetic Field}
\author{Prabal Adhikari}
\affiliation{Physics Department, Faculty of Natural Sciences and Mathematics, St. Olaf College, 1520 St. Olaf Avenue, Northfield, MN 55057, United States}
\begin{abstract}
\noindent
We study the topological susceptibility and the fourth cumulant of the QCD vacuum in the presence of a uniform, background magnetic field in two-and-three flavor QCD finding novel, model-independent sum rules relating the shift in the topological susceptibility due to the background field to the shift in the quark condensates, and the shift in the fourth cumulant to the shifts in the quark condensates and susceptibilities.
\end{abstract}
\maketitle 
\section{Introduction}
The vacuum of quantum chromodynamics (QCD) exhibits topological properties, which are characterized by CP-even topological cumulants. Their origin is intimately connected to the axial $U(1)$ problem~\cite{Crewther:1977ce,Witten:1979vv,Veneziano:1979ec,DiVecchia:1980yfw,SHIFMAN1980493}, namely that while the QCD Lagrangian possesses an axial $U(1)$ symmetry for massless quarks, a symmetry that is approximate for light quarks, but is not accompanied by parity doublets in the QCD spectrum or if spontaneously broken by an isosinglet pseudoscalar, which is also absent~\cite{Weinberg:1975ui}. The resolution to the axial $U(1)$ puzzle was suggested by Witten and Veneziano~\cite{Witten:1979vv,Veneziano:1979ec} based on the 't Hooft large-$N_{c}$ limit and the existence of instantons, which break the symmetry~\cite{tHooft:1976rip, tHooft:1976snw} with the proposed Witten-Veneziano formula confirmed on the lattice~\cite{DelDebbio:2004ns} using Ginsparg-Wilson fermions and a proof of the ultraviolet (uv) finiteness of the topological susceptibilty, which is non-trivial~\cite{GIUSTI2002234,GIUSTI2004157,LUSCHER2004296}. For a recent review of lattice studies, see Ref.~\cite{VICARI200993}.

While the QCD action is invariant under an axial rotation (for massless quarks), the integration measure associated with the quarks fields in the QCD partition function,
\begin{equation}
\begin{split}
\label{eq:ZQCD}
Z_{\rm QCD}=\int \mathcal{D}A\mathcal{D}q\mathcal{D}\bar{q}\exp\left [i\int d^{4}x\ \mathcal{L}_{\rm QCD} \right ]\ ,
\end{split}
\end{equation}
transforms~\cite{Fujikawa:1979ay,LUSCHER1998342}, giving rise to a non-conserved anomalous current. The QCD Lagrangian, $\mathcal{L}_{\rm QCD}$, with the $\theta$-term required to investigate the topological cumulants is 
\begin{equation}
\begin{split}
\label{eq:LQCD}
\mathcal{L}_{\rm QCD}=&-\frac{1}{4}G_{\mu\nu}^{a}G_{a}^{\mu\nu}-\frac{g^{2}\theta }{32\pi^{2}}\tilde{G}^{a}_{\mu\nu}G^{\mu\nu}_{a}\\
&+\bar{q}\left (i\slashed{D}-M \right )q\ ,
\end{split}
\end{equation}
where $G_{\mu\nu}^{a}$ is the gluon field strength tensor, $\tilde{G}^{a}_{\mu\nu}\equiv\frac{1}{2}\epsilon_{\mu\nu\alpha\beta}G^{\nu\alpha\beta}$ is the dual field strength tensor, and $q$ is the quark field with flavor and color indices suppressed. The covariant derivative that appears in $\slashed{D}\equiv\gamma^{\mu}D_{\mu}$ is $D_{\mu}=\partial_{\mu}-igt_{a}A^{a}_{\mu}$ with $A^{a}_{\mu}$ being the gluon field and $t^{a}$ the generator of the $SU(3)$ color gauge group. We will assume that the quark mass matrix, $M$, is real, which ensures that $\theta$ is a physical observable. It is constrained experimentally by the small size of the neutron dipole moment~\cite{Afach:2015sja} with a recent estimate of $\theta\lesssim10^{-11}$~\cite{kim2010axions}.

The QCD phase diagram in the presence of an external magnetic has drawn interest due to the relevance in magnetized neutron stars (magnetars), the quark-gluon plasma in the early universe and more recently in the context of heavy-ion collisions, where qualitative features of the chiral magnetic effect (CME) have been observed~\cite{2015:uf}. The focus of CME studies has been the role of chiral imbalance, introduced by an axial chemical potential (or a background electric field), which in the simultaneous presence of a (parallel) magnetic field generates an axial current proportional to both the axial chemical potential (or electric field) and the magnetic field~\cite{Kharzeev_2014}. However, the effect on the topological properties of the QCD vacuum due to the background magnetic field in the absence of an axial chemical potential within the confined phase of QCD has been completely (or perhaps largely) ignored. The objective of this paper is to fill this intellectual gap by characterizing the shifts in the topological susceptibility and the fourth cumulant due to the background magnetic field. In order to do so we upgrade the $SU(3)_{\rm color}$ gauge symmetry in the QCD Lagrangian to $SU(3)_{\rm color}\times U(1)_{\rm em}$,  which involves coupling the quark fields to an electromagnetic (em) gauge field, $A_{\mu}$, which is achieved by modifying the covariant derivative in Eq.~(\ref{eq:LQCD}), i.e. $\slashed{D}\rightarrow\slashed{D}-ieQ\slashed{A}$, where $Q$ is the quark charge matrix and including the standard quadratic contribution of the electromagnetic field.

Most recent analytical studies of the topological susceptibility and fourth cumulants have been conducted in the context of chiral perturbation theory ($\chi$PT)~\cite{Gasser:1983yg,Gasser:1984gg,scherer2011primer}, with the results valid for small quarks masses~\cite{Mao:2009sy,Guo:2015oxa,Bernard:2012ci} and large volumes~\cite{Leutwyler:1992yt}. Analytical studies prior to the development of $\chi$PT relied on current algebra techniques~\cite{SHIFMAN1980493}, with the results valid to linear order in the quark masses and $\theta$. In this letter, we study the first two topological cumulants of QCD in the presence of a magnetic field using $\chi$PT, with the results model-independent and valid for weak magnetic fields, i.e. $eH\ll (4\pi f_{\pi})^{2}$. Due to the non-perturbative nature of QCD, studies with larger quark masses and/or magnetic fields must be conducted on the lattice.

The paper is organized as follows: in the next two sections of this letter, we study two-flavor and three-flavor $\chi$PT, respectively, and find sum rules that relate the shifts (due to the background magnetic field) in the topological susceptibility and the fourth cumulant to the shifts in the quark condensates and susceptibilities. In the following section, we discuss a Ward-Takahashi identity for topological susceptibility valid in a background magnetic field and in the last section, we conclude and speculate on possible heavy ion and cosmological applications.

\section{Two-flavor $\chi$PT}
\label{sec:2f}
The $\mathcal{O}(p^{2})$ $\chi$PT Lagrangian relevant for $n$-flavor QCD at low energies is
\begin{equation}
\begin{split}
\label{eq:L2}
\mathcal{L}_{2}=&-\frac{1}{4}F_{\mu\nu}F^{\mu\nu}+\frac{f^{2}}{4}\Tr\left [\nabla_{\mu}\Sigma(\nabla^{\mu}\Sigma)^{\dagger} \right ]\\
&+\frac{f^{2}}{4}\Tr\left[\chi\Sigma^{\dagger}+\Sigma\chi^{\dagger} \right ]\ ,
\end{split}
\end{equation}
where $F_{\mu\nu}$ is the electromagnetic tensor, $\Sigma$ is an $SU(n)$ matrix, $f$ is the tree-level pion decay constant, $\chi$ is the scalar-pseudoscalar source, and $\nabla_{\mu}\Sigma=\partial_{\mu}\Sigma-ieA_{\mu}[Q,\Sigma]$ is the covariant derivative in the presence of an electromagnetic gauge field, $A_{\mu}$, with $Q$ being the quark charge matrix. The effects of the QCD $\theta$-term is fully incorporated into the rotated quark mass term via an axial rotation after which the scalar source term is $\chi=2Be^{-i\theta/n}M$. For instance, see Ref.~\cite{srednicki2007quantum} for an elementary discussion.

In two-flavor QCD, the mass matrix, $M={\rm diag}(m_{u},m_{d})$ points purely in the $\mathbb{1}$-direction in the isospin limit but has a component in the $\tau_{3}$-direction that is proportional to the difference in the quark masses. We, therefore, anticipate the possibility that the ground state, $\Sigma_{\alpha}$, has a non-zero component in the $\tau_{3}$-direction due to the $\theta$-term. In order to characterize this explicitly, we proceed by parameterizing the most general form for the ground state in the presence of the $\theta$ term,
\begin{align}
\Sigma_{\alpha}=\cos\alpha\ \mathbb{1}+i\sin\alpha\ \hat{\phi}_{a}\tau_{a}\ ,
\end{align}
where we use the Einstein summation convention, here and throughout, with an implied sum over the isospin index $a$ and $\hat{\phi}_{a}\hat{\phi}_{a}=1$ that ensures $\Sigma_{\alpha}$ is unitary. The  Lagrangian in the absence of a magnetic field is
\begin{equation}
\begin{split}
\mathcal{L}_{\rm tree}=&2f^{2}B\hat{m}\left[\cos\alpha\cos\tfrac{\theta}{2}+\hat{\phi}_{3}\delta\sin\alpha\sin\tfrac{\theta}{2}\right]\ ,
\end{split}
\end{equation}
where $\hat{m}=\tfrac{m_{u}+m_{d}}{2}$ and $\delta=\tfrac{m_{d}-m_{u}}{m_{u}+m_{d}}$, which is maximized for $\hat{\phi}_{3}=1$, assuming $m_{d}>m_{u}$. Consequently, $\hat{\phi}_{1}=\hat{\phi}_{2}=0$ and the tree-level Lagrangian is $\mathcal{L}_{\rm tree}=f^{2}B\left[m_{u}\cos\phi_{u}+m_{d}\cos\phi_{d}\right]$, with $\phi_{u}=\tfrac{\theta}{2}+\alpha$ and $\phi_{d}=\tfrac{\theta}{2}-\alpha$. Maximing $\mathcal{L}_{\rm tree}$ with respect to $\alpha$ we get 
\begin{equation}
\begin{split}
\label{eq:alpha-tree-2f}
\tan\alpha=\delta\tan\tfrac{\theta}{2}\ ,
\end{split}
\end{equation}
which is non-zero when $\delta$ and $\theta$ are both non-zero.

In order to study the changes in the topological fluctuations of the QCD vacuum in a background magnetic field, a next-to-leading order (NLO) effect, we proceed by parameterizing the fluctuations of the pion fields in order to incorporate the interactions of the charged meson pairs with the external field, $\Sigma=e^{i\frac{\alpha}{2}\tau_{3}}e^{i\frac{\phi_{a}\tau_{a}}{f}}e^{i\frac{\alpha}{2}\tau_{3}}$, where $\phi_{a}$ with $a=1,2,3$ represent field fluctuations. Using the standard definition of charge eigenstates~\cite{scherer2011primer},
the quadratic Lagrangian is
\begin{align}
\label{eq:Lquad-2f}
\nonumber
\mathcal{L}_{\rm quad}&=\frac{1}{2}H^{2}+D_{\mu}\pi^{+}D^{\mu}\pi^{-}-\mathring{m}_{\pi}^{2}(\theta)\pi^{+}\pi^{-}\\
&+\frac{1}{2}\partial_{\mu}\pi^{0}\partial^{\mu}\pi^{0}-\frac{1}{2}\mathring{m}_{\pi}^{2}(\theta)\pi^{0}\pi^{0}\ ,
\end{align}
where $\mathring{m}_{\pi}^{2}(\theta)$ is the degenerate tree-level pion mass in the $\theta$-vacuum, 
\begin{align}
\label{eq:Sigmags}
\mathring{m}_{\pi}^{2}(\theta)&=B\left [m_{u}\cos\phi_{u}(\theta)+m_{d}\cos\phi_{d}(\theta) \right ]\ ,
\end{align}
and $D_{\mu}\pi^{\pm}=(\partial_{\mu}\pm ieA_{\mu})\pi^{\pm}$ are the covariant derivatives associated with the $U(1)$ gauge field. For an $\mathcal{O}(p^{4})$ calculation that incorporates the effects of a background magnetic field, we require one loop contributions arising from $\mathcal{L}_{\rm quad}$ and the tree-level counterterms from the $\mathcal{O}(p^{4})$ $\chi$PT Lagrangian~\cite{scherer2011primer}. Using the the ground state orientation, $\Sigma_{\alpha}=e^{i\alpha\tau_{3}}$, which follows from Eq.~(\ref{eq:Sigmags}) with $\hat{\phi}_{3}=1$ and consistent with the parameterization of $\Sigma$ above Eq.~(\ref{eq:Lquad-2f}), we get the tree-level counterterm
\begin{equation}
\begin{split}
\mathcal{L}_{{\rm ct}}=&(l_{3}+l_{4})\left [B\left\{m_{u}\cos\phi_{u}(\theta)+m_{d}\cos\phi_{d}(\theta)\right\} \right ]^{2}\\
+&l_{7}\left [B\left\{m_{u}\sin\phi_{u}(\theta)+m_{d}\sin\phi_{d}(\theta)\right\}\right ]^{2}\\
+&(h_{1}+h_{3}-l_{4})\left[B^{2}(m_{u}^{2}+m_{d}^{2})\right]-4h_{2}(eH)^{2}\\
+&2(h_{1}-h_{3}-l_{4})\left[B^{2}m_{u}m_{d}\cos(\theta)\right]\ .
\end{split}
\end{equation}
The low ($l_{i}$) and high ($h_{i}$) energy constants required for renormalization are
\begin{align}
\label{eq:lihi}
l_{i}&=l_{i}^{r}+\gamma_{i}\lambda\ ,\ \ h_{i}=h_{i}^{r}+\delta_{i}\lambda\ ,
\end{align}
where $\lambda=-\frac{\Lambda^{-2\epsilon}}{2(4\pi)^{2}}\left(\frac{1}{\epsilon}+1\right)$, $l_{i}^{r}$ and $h_{i}^{r}$ are depend on the $\overline{\rm MS}$ renormalization scale $\Lambda$, and the constants $\gamma_{i}$ and $\delta_{i}$ are
$\gamma_{4}=\delta_{1}=2$, $\gamma_{5}=-\tfrac{1}{6}$, $\gamma_{7}=\delta_{3}=0$ and $\delta_{2}=\tfrac{1}{12}$.
The one-loop contribution of the charged pions requires summing over all the Landau levels, which is most conveniently done using the dimensionally regularized Schwinger proper time integral,
\begin{equation}
\begin{split}
I_{H}[\mathring{m}_{\pi}(\theta)]&=-\frac{\mu^{2\epsilon}}{(4\pi)^{2}}\int_{0}^{\infty} \frac{ds}{s^{3-\epsilon}}e^{-\mathring{m}^{2}_{\pi}(\theta)s}\frac{eHs}{\sinh eHs}
\end{split}
\end{equation}
with $\mu^{2}=e^{\gamma_{E}}\Lambda^{2}$. The integral is uv divergent: for small values of $s$, the integrand has an $H$-independent, $s^{-3}$ divergence and an $s^{-1}$ divergence that is quadratic in the background magnetic field. Isolating both the $\mathring{m}_{\pi}(\theta)$ and $H$-dependent divergences we get,
\begin{align}
I_{H}[\mathring{m}_{\pi}(\theta)]=&I_{H}^{\rm div}[\mathring{m}_{\pi}(\theta)]+I_{H}^{\rm fin}[\mathring{m}_{\pi}(\theta)]\\
\nonumber
I_{H}^{\rm div}[\mathring{m}_{\pi}(\theta)]=&-\frac{\mathring{m}_{\pi}^{4}(\theta)}{2(4\pi)^{2}}\left[\frac{1}{\epsilon}+\frac{3}{2}+\log\frac{\Lambda^{2}}{\mathring{m}_{\pi}^{2}(\theta)} \right]\\
&+\frac{(eH)^{2}}{6(4\pi)^{2}}\left[\frac{1}{\epsilon}+\log\frac{\Lambda^{2}}{\mathring{m}_{\pi}^{2}(\theta)}\right]\\
I_{H}^{\rm fin}[\mathring{m}_{\pi}(\theta)]=&\frac{(eH)^{2}}{(4\pi)^{2}}\mathfrak{I}_{H}(\tfrac{\mathring{m}_{\pi}^{2}(\theta)}{eH})\ ,
\end{align}
where $\mathfrak{I}_{H}(y)=-\int_{0}^{\infty} \frac{dz}{z^{3}}e^{-yz}\left(\frac{z}{\sinh z}-1+\frac{z^{2}}{6}\right)$ and the finite contribution is in agreement with Ref.~\cite{Schwinger:1951nm}. 
Adding this contribution to the one-loop free energy of the neutral pion, the tree-level counter-term and the background magnetic field, the divergences cancel resulting in the one-loop free energy, $\mathcal{F}(\theta,H)\equiv\mathcal{F}_{0}(\theta)+\mathcal{F}_{H}(\theta)$, where $\mathcal{F}_{0}(\theta)$ is the $H$-independent contribution in the presence of a $\theta$ vacuum, 
\begin{align}
\label{F0-2f}
\nonumber
&\mathcal{F}_{0}(\theta)=-f^{2}\mathring{m}_{\pi}^{4}(\theta)-(l^{r}_{3}+h^{r}_{1})\mathring{m}_{\pi}^{4}(\theta)\\
\nonumber
&-\frac{3\mathring{m}_{\pi}^{4}(\theta)}{4(4\pi)^{2}}\left [\frac{1}{2}+\log\frac{\Lambda^{2}}{\mathring{m}_{\pi}^{2}(\theta)} \right]\\
\nonumber
&-h_{3}B^{2}\{m_{u}^{2}+m_{d}^{2}-2m_{u}m_{d}\cos\theta\}\\
&-l_{7}\left [B\left\{m_{u}\sin\phi_{u}(\theta)+m_{d}\sin\phi_{d}(\theta)\right\}\right ]^{2}\ ,
\end{align}
while $\mathcal{F}_{H}(\theta)$ is $H$-dependent,
\begin{equation}
\begin{split}
\label{eq:FH-2f}
\mathcal{F}_{H}(\theta)=\frac{1}{2}H_{R}^{2}+\frac{(eH)^{2}}{(4\pi)^{2}}\mathfrak{I}_{H}(\tfrac{\mathring{m}_{\pi}^{2}(\theta)}{eH})\ ,
\\
\end{split}
\end{equation}
where $H_{R}=Z_{e}^{-1}H$ is renormalized magnetic field and $Z_{e}$ is the charge renormalization wave function,
\begin{equation}
\begin{split}
Z_{e}&=1-4e^{2}h^{r}_{2}-\frac{e^{2}}{6(4\pi)^{2}}\left(\log\frac{\Lambda^{2}}{\mathring{m}_{\pi}^{2}(\theta)}-1\right)\ ,
\end{split}
\end{equation}
that keeps the product $eH$ unaltered, i.e. $e_{R}H_{R}=eH$. While the vacuum orientation, $\Sigma_{\alpha}$, is altered by NLO effects, the free energy is independent of the change in orientation at $\mathcal{O}(p^{4})$~\cite{Guo:2015oxa}. Consequently we can use the tree-level ground state of Eq.~(\ref{eq:alpha-tree-2f}) in order to calculate the shifts in the topological susceptibility and fourth cumulant. They are defined as
\begin{equation}
\begin{split}
\label{eq:cumulant-def}
\chi_{t,H}&=\left.\frac{\partial^{2}\mathcal{F}_{H}}{\partial\theta^{2}}\right|_{\theta=0},\ c_{4,H}=\left.\frac{\partial^{4}\mathcal{F}_{H}}{\partial\theta^{4}}\right|_{\theta=0}\ ,
\end{split}
\end{equation}
where $\mathcal{F}_{H}$ is the shift in the free energy due to the background magnetic field. Using analogous definitions for $H=0$, we reproduce the results for the NLO topological susceptibility and fourth cumulant of Ref.~\cite{Guo:2015oxa}. On the other hand, the shift of the topological susceptibility due to the background magnetic field is
\begin{equation}
\begin{split}
\label{eq:chitH}
\chi_{t,H}&=-\frac{B\bar{m}(eH)}{(4\pi)^{2}}\mathcal{I}_{H,2}(\tfrac{\mathring{m}_{\pi}^{2}}{eH})\ ,
\end{split}
\end{equation}
where $\mathring{m}_{\pi}^{2}\equiv \mathring{m}_{\pi}^{2}(0)$ is the degenerate pion mass in the $\theta=0$ vacuum, the integral $\mathcal{I}_{H,n}(y)$ is defined as $\mathcal{I}_{H,n}(y)=\int_{0}^{\infty}\frac{dz}{z^{n}}e^{-yz}\left(\frac{z}{\sinh z}-1\right)$ and the contributions purely quadratic in the magnetic field cancel. Similarly, the shift in the fourth cumulant is
\begin{equation}
\begin{split}
\label{eq:c4H}
c_{4,H}=&\frac{B\bar{m}^{4}(eH)}{(4\pi)^{2}}\left(\frac{1}{m_{u}^{3}}+\frac{1}{m_{d}^{3}}\right)\mathcal{I}_{H,2}(\tfrac{\mathring{m}_{\pi}^{2}}{eH})\\
-&\frac{3B^{2}\bar{m}^{2}}{(4\pi)^{2}}\mathcal{I}_{H,1}(\tfrac{\mathring{m}_{\pi}^{2}}{eH})\ ,
\end{split}
\end{equation}
where $\bar{m}=(\sum_{q_{f}}\frac{1}{m_{q_{f}}})^{-1}$ is the two-flavor reduced quark mass and $\mathring{m}_{\pi}\equiv\mathring{m}_{\pi}(0)$. Using the definitions of the quark condensates and susceptibilities, $\langle\bar{q}_{f}q_{f}\rangle_{H}=\frac{\partial \mathcal{F}_{H}}{\partial m_{q_{f}}}$ and $\chi_{q_{f},H}=\frac{\partial^{2} \mathcal{F}_{H}}{\partial m_{q_{f}}^{2}}$, where $q_{f}$ represents a particular quark flavor, we find that their shifts due to the background magnetic field are flavor degenerate,
\begin{align}
\label{eq:qqH}
\langle\bar{u}u\rangle_{H}&=\langle\bar{d}d\rangle_{H}=\frac{B(eH)}{(4\pi)^{2}}\mathcal{I}_{H,2}(\tfrac{\mathring{m}_{\pi}^{2}}{eH})\\
\label{eq:chiqH}
\chi_{u,H}&=\chi_{d,H}=-\frac{B^{2}}{(4\pi)^{2}}\mathcal{I}_{H,1}(\tfrac{\mathring{m}_{\pi}^{2}}{eH})\ .
\end{align}
Then using Eqs.~(\ref{eq:chitH}) and (\ref{eq:qqH}), we find the following two-flavor sum rule relating the shift in the topological susceptibility to the shift in the light quark condensates
\begin{equation}
\begin{split}
\chi_{t,H}&=-\bar{m}\langle\bar{q}_{f}q_{f}\rangle_{H}\ ,
\end{split}
\end{equation}
where $q_{f}$ is either $u$ or $d$. Using Eqs.~(\ref{eq:c4H}), (\ref{eq:qqH}) and (\ref{eq:chiqH}), we find the following two-flavor sum rule relating the shift in the fourth cumulant to the shifts in the quark condensates and quark susceptibilities, can be further simplified
\begin{equation}
\begin{split}
c_{4,H}&=\bar{m}^{4}\bigg(\sum_{q_{f}=u,d}\frac{\langle\bar{q}_{f}q_{f}\rangle_{H}}{m_{q_{f}}^{3}}\bigg)+3\bar{m}^{2}\chi_{q_{f},H}\ ,
\end{split}
\end{equation}
where $\bar{m}$ is the (two-flavor) reduced mass and $m_{q_{f}}$ are the quark masses. In the first term there is a sum over quark flavors, while in the second term $q_{f}$ is either $u$ or $d$.
The shift in the topological susceptibility is proportional to both the reduced mass and the chiral condensate. Since the magnetic field increases the magnitude of the quark condensate -- it becomes more negative --- the topological susceptibility is enhanced. On the other hand, the fourth cumulant shift is proportional to both the quark condensates and the quark susceptibilities. Since the terms have opposite signs, the fourth cumulant suppressed at weak fields and enhanced at strong fields with a non-zero critical magnetic field for which the fourth cumulant remains unaltered. 

\section{Three-flavor $\chi$PT}
\label{sec:3f}
We now turn our attention to three-flavor QCD~\cite{Gasser:1984gg}, the analysis of which follows similarly to two-flavor QCD with the scalar source, $\chi=2B_{0}{\rm diag}(m_{u},m_{d},m_{s})e^{i\theta/3}$ and $\Sigma=\Sigma_{\alpha_{i}}^{1/2}\exp\left(\tfrac{i\phi_{a}\lambda_{a}}{f}\right)\Sigma^{1/2}_{\alpha_{i}}$, where $\lambda_{a}$ are the Gell-Mann matrices, $\phi_{a}$ are the field fluctuations and $\Sigma_{\alpha_{i}}={\rm diag}(e^{i\alpha_{1}},e^{i\alpha_{2}},e^{i\alpha_{3}})$. Using the $\mathcal{O}(p^{2})$ Lagrangian of Eq.~(\ref{eq:L2}) we get the tree-level Lagrangian with $\sum_{i}\alpha_{i}=1$,
\begin{equation}
\begin{split}
\mathcal{L}_{\rm tree}&=f^{2}B_{0}\sum_{q_{f}=u,d,s}m_{q_{f}}\cos\phi_{q_{f}}(\theta)\ ,
\end{split}
\end{equation}
where $\phi_{u}=\tfrac{\theta}{3}-\alpha_{1}$, $\phi_{d}=\tfrac{\theta}{3}-\alpha_{2}$, $\phi_{s}=\tfrac{\theta}{3}-\alpha_{3}$ and we have ignored the tree-level contribution of the external magnetic field. In the three-flavor case, the tree-level potential can be minimized recursively~\cite{LU_2020} for arbitrary values of $\theta$, which gives,
\begin{equation}
\begin{split}
\phi_{q_{f}}&=\frac{\bar{m}}{m_{q_{f}}}\theta+\frac{\bar{m}}{m_{q_{f}}}\left(\frac{\bar{m}^{2}}{m_{q_{f}}^{2}}-\sum_{\tilde{q}_{f}}\frac{\bar{m}^{3}}{m_{\tilde{q}_{f}}^{3}}\right)\frac{\theta^{3}}{6}+\dots\ ,
\end{split}
\end{equation}
where $\bar{m}$ is now the three-flavor reduced quark mass, $q_{f}$ is either $u$, $d$ or $s$, and the dots represent higher order corrections beginning at $\mathcal{O}(\theta^{5})$.
Similar to the two-flavor case, quantum fluctuations alter this ground state but do not affect the $\mathcal{O}(p^{4})$ free energy~\cite{Guo:2015oxa}. As before, we require the contributions of the quadratic Lagrangian, similar to Eq.~(\ref{eq:Lquad-2f}) but with contributions from the meson octet. Since we are only interested in the shifts induced by the background magnetic field, the masses that we require are those of the charged pions and the charged kaons,
\begin{align}
\mathring{m}_{\pi^{\pm}}^{2}(\theta)&=B_{0}[m_{u}\cos\phi_{u}(\theta)+m_{d}\cos\phi_{d}(\theta)]\\
\mathring{m}_{K^{\pm}}^{2}(\theta)&=B_{0}[m_{u}\cos\phi_{u}(\theta)+m_{s}\cos\phi_{s}(\theta)]\ .
\end{align}
The one-loop contribution to the free energy required for an NLO calculation consists of divergences that are canceled by the divergence in the tree-level counterterm of the same order, i.e. $\mathcal{O}(p^{4})$,
\begin{align}
\nonumber
\mathcal{L}_{\rm ct}&=16L_{6}B_{0}^{2}\Bigg[\sum_{q_{f}=u,d,s}m_{q_{f}}\cos\phi_{q_{f}}(\theta)\Bigg]^{2}\\
\nonumber
&-16L_{7}B_{0}^{2}\Bigg[\sum_{q_{f}=u,d,s}m_{q_{f}}\sin\phi_{q_{f}}(\theta)\Bigg]^{2}\\
\nonumber
&+8L_{8}B_{0}^{2}\sum_{q_{f}=u,d,s}m_{q_{f}}^{2}\cos[2\phi_{q_{f}}(\theta)]\\
&+4H_{2}B_{0}^{2}\sum_{q_{f}=u,d,s}m_{q_{f}}^{2}+\frac{4}{3}(L_{10}+2H_{1})(eH)^{2}\ ,
\end{align}
where $L_{i}$ with $i=1,2\dots 8$ and $H_{2}$ are the relevant low and high energy constants defined analogously to Eq. (\ref{eq:lihi}) with $\delta_{i}$ replaced by $\Delta_{i}$ and $\gamma_{i}$ replaced by $\Gamma_{i}$, where $\Gamma_{6}=\tfrac{11}{144}$, $\Gamma_{7}=0$, $\Gamma_{8}=\tfrac{5}{48}$, $\Gamma_{10}=-\tfrac{1}{4}$, $\Delta_{1}=-\tfrac{1}{8}$ and $\Delta_{2}=\tfrac{5}{24}$.
Combining the tree-level and one-loop contributions, we get the following three-flavor free energy in the presence of the $\theta$-vacuum,
\begin{align}
\label{eq:F-3f}
\nonumber
&\mathcal{F}(\theta)=\frac{1}{2}H_{R}^{2}-f^{2}B_{0}\sum_{q_{f}}m_{q_{f}}\cos\phi_{q_{f}}(\theta)\\
\nonumber
&-\sum_{i}\frac{\mathring{m}_{i}(\theta)^{4}}{4(4\pi)^{2}}\left [\frac{1}{2}+\log\frac{\Lambda^{2}}{\mathring{m}_{i}(\theta)^{2}}  \right ]\\
\nonumber
&-16L^{r}_{6}B_{0}^{2}\Big[\sum_{q_{f}}m_{q_{f}}\cos\phi_{q_{f}}(\theta)\Big]^{2}\\
\nonumber
&+16L^{r}_{7}B_{0}^{2}\Big[\sum_{q_{f}}m_{q_{f}}\sin\phi_{q_{f}}(\theta)\Big]^{2}\\
\nonumber
&-8L^{r}_{8}B_{0}^{2}\sum_{q_{f}}m_{q_{f}}^{2}\cos[2\phi_{q_{f}}(\theta)]-4H^{r}_{2}B_{0}^{2}\sum m_{q_{f}}^{2}\\
&+\frac{(eH)^{2}}{(4\pi)^{2}}\left[\mathfrak{I}_{H}(\tfrac{\mathring{m}_{\pi^{\pm}}(\theta)^{2}}{eH})+\mathfrak{I}_{H}(\tfrac{\mathring{m}_{K^{\pm}}(\theta)^{2}}{eH})\right]\ ,
\end{align}
where the sum in $i$ is over the meson octet and the sum in $q_{f}$ is over the three quark flavors, and  $H_{R}=Z_{e}^{-1}H$ is the renormalized magnetic field with $Z_{e}$ being the charge renormalization factor,
\begin{align}
Z_{e}=1&+\frac{4e^{2}}{3}\bigg[\ell^{r}-\frac{1}{4(4\pi)^{2}}\bigg(\sum_{c=\pi,K}\log\tfrac{\Lambda^{2}}{\mathring{m}^{2}_{c^{\pm}}}-2\bigg)\bigg]
\end{align}
where $\ell^{r}\equiv L^{r}_{10}+2H^{r}_{1}$. The topological susceptibility and the fourth cumulant shifts are then calculated using the $H$-dependent contributions to the free energy defined in Eq.~(\ref{eq:F-3f}) and the definitions of the cumulant shifts in Eqs.~(\ref{eq:cumulant-def}). We get the following topological susceptibility shift
\begin{equation}
\begin{split}
\label{eq:chitH-3f}
\chi_{t,H}=&-\frac{B_{0}\bar{m}^{2}(eH)}{(4\pi)^{2}}\left[\left(\frac{1}{m_{u}}+\frac{1}{m_{d}}\right) \mathcal{I}_{H,2}(\tfrac{\mathring{m}^{2}_{\pi^{\pm}}}{eH})\right.\\
&\left.+\left(\frac{1}{m_{u}}+\frac{1}{m_{s}}\right)\mathcal{I}_{H,2}(\tfrac{\mathring{m}^{2}_{K^{\pm}}}{eH})\right]\ ,
\end{split}
\end{equation}
and the following fourth cumulant shift
\begin{equation}
\begin{split}
\label{eq:c4H-3f}
c_{4,H}&=\frac{B_{0}\bar{m}^{4}(eH)}{(4\pi)^{2}}\left[\left(\frac{1}{m_{u}^{3}}+\frac{1}{m_{d}^{3}}\right)\mathcal{I}_{H,2}(\tfrac{\mathring{m}^{2}_{\pi^{\pm}}}{eH})\right.\\
&\left.+\left(\frac{1}{m_{u}^{3}}+\frac{1}{m_{s}^{3}}\right)\mathcal{I}_{H,2}(\tfrac{\mathring{m}^{2}_{K^{\pm}}}{eH})\right]\\
&-\frac{3B_{0}^{2}\bar{m}^{5}}{(4\pi)^{2}}\left[\frac{1}{\bar{m}_{ud}}\left(\frac{1}{m_{u}}+\frac{1}{m_{d}}\right)^{2}\mathcal{I}_{H,1}(\tfrac{\mathring{m}^{2}_{\pi^{\pm}}}{eH})\right.\\
&\left.+\frac{1}{\bar{m}_{us}}\left(\frac{1}{m_{u}}+\frac{1}{m_{s}}\right)^{2}\mathcal{I}_{H,1}(\tfrac{\mathring{m}^{2}_{K^{\pm}}}{eH})\right]\ ,
\end{split}
\end{equation}
where $\mathring{m}_{\pi^{\pm}}\equiv \mathring{m}_{\pi^{\pm}}(0)$ and $\mathring{m}_{K^{\pm}}\equiv\mathring{m}_{K^{\pm}}(0)$. The mass $\bar{m}_{uq_{f}}$ is
\begin{align}
\bar{m}_{uq_{f}}^{-1}&=\frac{1}{m_{u}}+\frac{1}{m_{q_{f}}}-\frac{3}{m_{\tilde{q}_{f}}}+\tfrac{\frac{4}{m_{\tilde{q}_{f}}^{3}}}{\left(\frac{1}{m_{u}^{2}}+\frac{1}{m_{q_{f}}^{2}}-\frac{1}{m_{u}m_{q_{f}}}\right)}\ ,
\end{align}
where $q_{f}=d$ and $\tilde{q}_{f}=s$ or $q_{f}=s$ and $\tilde{q}_{f}=d$. In the large $m_{s}$ limit, the terms containing the kaon masses are exponential suppressed and $\bar{m}_{ud}$ reduces to the two-flavor reduced mass, thus reproducing results in the two-flavor case.

In order to relate these to shifts in the quark condensates and susceptibilities, we first note that unlike the two-flavor case, the shifts of the quark condensates are non-degenerate,
\begin{align}
\langle \bar{u}u\rangle_{H}&=\frac{B_{0}eH}{(4\pi)^{2}}\left[\mathcal{I}_{H,2}(\tfrac{\mathring{m}^{2}_{\pi^{\pm}}}{eH})+\mathcal{I}_{H,2}(\tfrac{\mathring{m}^{2}_{K^{\pm}}}{eH})\right]\\
\langle \bar{d}d\rangle_{H}&=\frac{B_{0}eH}{(4\pi)^{2}}\mathcal{I}_{H,2}(\tfrac{\mathring{m}^{2}_{\pi^{\pm}}}{eH})\\
\langle \bar{s}s\rangle_{H}&=\frac{B_{0}eH}{(4\pi)^{2}}\mathcal{I}_{H,2}(\tfrac{\mathring{m}^{2}_{K^{\pm}}}{eH})\ ,
\end{align}
which is also true for the shifts of the quark susceptibilities,
\begin{align}
\chi_{u,H}&=-\frac{B_{0}^{2}}{(4\pi)^{2}}\left[\mathcal{I}_{H,1}(\tfrac{\mathring{m}^{2}_{\pi^{\pm}}}{eH})+\mathcal{I}_{H,1}(\tfrac{\mathring{m}^{2}_{K^{\pm}}}{eH})\right]\\
\chi_{d,H}&=-\frac{B_{0}^{2}}{(4\pi)^{2}}\mathcal{I}_{H,1}(\tfrac{\mathring{m}^{2}_{\pi^{\pm}}}{eH})\\
\chi_{s,H}&=-\frac{B_{0}^{2}}{(4\pi)^{2}}\mathcal{I}_{H,1}(\tfrac{\mathring{m}^{2}_{K^{\pm}}}{eH})\ .
\end{align}
These shifts satisfy the following sum rules,
\begin{align}
\label{eq:sumruleqq-3f}
\langle\bar{u}{u}\rangle_{H}&=\langle\bar{d}{d}\rangle_{H}+\langle\bar{s}s\rangle_{H}\\
\label{eq:sumrulechiq-3f}
\chi_{u,H}&=\chi_{d,H}+\chi_{s,H}\ ,
\end{align}
the origin of which is explained by the valence quarks and anti-quarks in the charge eigenstates, $\pi^{\pm}$ and $K^{\pm}$. Combining the first of these sum rules with the shift in the topological susceptibility of Eq.~(\ref{eq:chitH-3f}), we find that the topological susceptibility shift is related to the quark condensate shifts via the three-flavor sum rule,
\begin{align}
\label{eq:sumrulechit-3f}
\chi_{t,H}=&-\bar{m}^{2}\sum_{q_{f}=u,d,s}\frac{\langle\bar{q}_{f}q_{f}\rangle_{H} }{m_{q_{f}}}\ ,
\end{align}
where $\bar{m}$ is the three-flavor reduced mass. Similarly, using the fourth cumulant shift in Eq.~(\ref{eq:c4H-3f}) and the quark condensate sum rule of Eq. (\ref{eq:sumruleqq-3f}), we get the three-flavor sum rule for the fourth cumulant shift,
\begin{align}
\nonumber
c_{4,H}=&\bar{m}^{4}\sum_{q_{f}=u,d,s}\frac{\langle\bar{q}_{f}q_{f}\rangle_{H}}{m_{q_{f}}^{3}}+3\bar{m}^{5}\left[\frac{1}{\bar{m}_{ud}}\left(\frac{1}{m_{u}}+\frac{1}{m_{d}}\right)^{2}\chi_{d,H}\right.\\
\label{eq:sumrulechit-3f}
&\left.+\frac{1}{\bar{m}_{us}}\left(\frac{1}{m_{u}}+\frac{1}{m_{s}}\right)^{2}\chi_{s,H}\right]\ ,
\end{align}
which can be recast in terms of $\chi_{u,H}$ using the quark condensate sum rule of Eq. (\ref{eq:sumrulechiq-3f}) at the expense of either $\chi_{d,H}$ or $\chi_{s,H}$. In the large $m_{s}$ limit, the integrals involving kaons are exponentially suppressed and $B_{0}$ can be identified with $B$ at leading order~\cite{Gasser:1984gg} with the two-flavor sum rules reducing to two-flavor ones. Similar to the two-flavor case, the topological susceptibility, according to Eq.~(\ref{eq:sumrulechit-3f}), is enhanced. However, the shift is now proportional to the reduced mass squared instead of the reduced mass, which is a consequence of the fact that up-and-down quark condensate shifts are non-degenerate. 

\section{Ward-Takahashi Identity for Topological Susceptibility}
The sum rules that we have discussed in the context of $\chi$PT for the topological susceptibility and the fourth cumulant are strictly valid at weak magnetic fields. However, it is possible to construct Ward-Takahashi (WT) identities valid in QCD for arbitrary masses and magnetic fields using the QCD partition function, $Z_{\rm QCD}$, of Eq.~(\ref{eq:ZQCD}), without calculating the individual correlation functions, which cannot be done analytically due to the non-perturbative nature of QCD. The WT identity is particularly insightful in the context of the topological susceptibility -- the finite $H$ sum rule assuming degenerate quark masses according to $\chi$PT is 
\begin{align}
\label{eq:topiso}
\chi_{t,H}&=-m_{q_{f}}\frac{\langle \bar{q}_{f}q_{f}\rangle_{H}}{n}\ ,
\end{align}
where $n=2$ or $3$, $m_{q_{f}}$ is the degenerate quark mass and $\langle \bar{q_{f}}q_{f}\rangle_{H}$ is the degenerate quark condensate shift in a background magnetic field with $q_{f}$ equal to either $u$ or $d$ or $s$ with the latter excluded in the two-flavor case. It has a structure similar to that of the $H=0$ tree-level relation between the topological susceptibility and the quark condensate. A closer examination suggests that this identity is a manifestation of a WT identity for the topological susceptibility, which relates it the quark condensates and a two-point correlation function of the singlet, pseudoscalar (CP-odd) operators, $\bar{q}_{f}\gamma_{5}q_{f}$. In order to derive this, one can proceed by promoting the $\theta$-angle to have spacetime dependence, i.e. $\theta\rightarrow\theta(x)$, and taking two functional derivatives with respect to $\theta(x)$ of the partition function using the QCD Lagrangian in Eq.~(\ref{eq:LQCD}) and with the QCD Lagrangian with all the $\theta$-dependence rotated into the quark mass term. Since using the two Lagrangians is equivalent, we find after integrating over one of the space-time variables and separating out the $H$-dependent contribution, the following Ward-Takahashi identity,
\begin{align}
\nonumber
\chi_{t,H}&=-\frac{1}{n^{2}}\langle\bar{q}Mq\rangle_{H}\\
&+\frac{i}{n^{2}}\int_{x}\langle \mathcal{T}\bar{q}(x)M\gamma_{5}q(x)\bar{q}(0)M\gamma_{5}q(0)\rangle_{H}\ ,
\end{align}
where $M$ is the quark mass matrix, $\mathcal{T}$ denotes time-ordering, $\int_{x}\equiv \int d^{4}x$, and we have suppressed all color and flavor indices. For degenerate quark masses, we note that the result is consistent with Eq.~(\ref{eq:topiso}) and $\chi$PT requires that the $\mathcal{O}(M^{2})$ contribution of the integrated two-point correlation function of $\bar{q}_{f}\gamma_{5}q_{f}$ vanish. For non-degenerate quark masses, we can use the $\chi$PT sum rules to determine the integrated two-point correlation function. We get, in the two-flavor case,
\begin{align}
\nonumber
&i\int_{x}\langle \mathcal{T}\bar{q}(x)M\gamma_{5}q(x)\bar{q}(0)M\gamma_{5}q(0)\rangle_{H}\\
=&\frac{m_{d}-m_{u}}{m_{u}+m_{d}}\left[-m_{u}\langle\bar{u}u\rangle_{H}+m_{d}\langle\bar{d}d\rangle_{H}\right]\ .
\end{align}
and in the three-flavor case,
\begin{align}
\nonumber
&i\int_{x}\langle \mathcal{T}\bar{q}(x)M\gamma_{5}q(x)\bar{q}(0)M\gamma_{5}q(0)\rangle_{H}\\
\nonumber
=&m_{u}\left[1-\frac{m_{d}^{2}m_{s}^{2}}{[\tfrac{1}{3}(m_{u}m_{d}+m_{d}m_{s}+m_{s}m_{u})]^{2}}\right]\langle\bar{u}u\rangle_{H}\\
&+{\rm (cylic\ permutations)}\ .
\end{align}
In the two-flavor case, the integrated two-point correlation function is proportional to the quark mass difference, which is not the case in the three-flavor case. Nevertheless, they both vanish for degenerate quark masses.
\section{Conclusion}
\label{sec:conclusion}
In this letter, we have generalized model-independent studies of topological cumulants to include the effect of a background magnetic field. The intimate relationship between topological susceptibility and the chiral condensate has been known since seminal studies conducted in the seventies and eighties~\cite{Crewther:1977ce,Witten:1979vv,Veneziano:1979ec,DiVecchia:1980yfw,Gasser:1983yg,Gasser:1984gg} and in this letter, we have characterized in a model-independent manner the connection between the shift in the topological susceptibility due to a background magnetic field to that of the chiral condensate and found that the shifts are proportional to each of the quark condensate shifts. We also find that this proportionality is a general feature of QCD valid for all quark masses and magnetic fields. While the topological susceptibility is enhanced for all magnetic fields within $\chi$PT, the fourth cumulant exhibits both suppression and enhancement at weak and strong fields respectively, with a zero shift at a critical, non-zero magnetic field. Determining the precise value of this critical magnetic field will require a careful study involving physical parameters as does the precise shift in the topological susceptibility. These will be pursued in forthcoming studies~\cite{Adhikari:2021xra}.

While the WT identity discussed here holds generally in QCD, the $\chi$PT sum rules derived here, while model-independent, are strictly speaking valid at only at small quark masses and weak magnetic fields though there are no constraints on the value of $\theta$ itself since it appears as a phase in the scalar source term of $\chi$PT. Due to the absence of the fermion sign problem in the simultaneous presence of $\theta$ and an external magnetic field, it should be possible to test the validity of these sum rules not only in the regime of validity of $\chi$PT but also for intermediate and large quark masses and/or magnetic fields. Furthermore, granted magnetic fields and/or rotational dynamics are important in heavy ion collisions, the effect on topological cumulants may be studied experimentally. Finally, there may also be cosmological and astrophysical applications given the presence of magnetic fields in the early universe and magnetars.
\section{Acknowledgements}P.A. would like to acknowledge St. Olaf College startup funds that partially supported this work. P.A. is also indebted to the referees for their illuminating insights and suggestions. 
\bibliographystyle{apsrev4-1}
\bibliography{/Users/prabal7e7/Documents/Research/bib}

\end{document}